\documentclass[a4paper, 12pt]{article}
\usepackage{geometry}
\usepackage[margin=.8cm]{caption}
\usepackage[utf8]{luainputenc}
\usepackage{times}
\usepackage{mathtools,amssymb}
\usepackage{microtype}
\usepackage[hidelinks]{hyperref}
\usepackage{units}
\usepackage{enumitem}
\usepackage{marginnote}
\usepackage[table]{xcolor}
\usepackage{pgfplots}
\pgfplotsset{every tick label/.append style={font=\scriptsize}}
\usepackage{makron}
\usepackage{graphicx}

\graphicspath{{./illustrationer/}}

\title{Out-of-Band Radiation from \mbox{Large Antenna Arrays}}
\author{Christopher Mollén, Erik G. Larsson, Ulf Gustavsson,\\ Thomas Eriksson and Robert W. Heath, Jr.}

\newcommand{\copyrightnotice}{\small This work has been submitted to the IEEE for possible publication. Copyright may be transferred without notice, after which this version may no longer be accessible.}

\makeatletter
\def\ps@firstpage{%
   	\let\@oddhead\@empty%
   	\let\@evenhead\@empty%
   	\def\@oddfoot{\fbox{\parbox{\linewidth}{\copyrightnotice}}}%
   	\def\@evenfoot{\fbox{\parbox{\linewidth}{\copyrightnotice}}}%
}
\makeatother

\begin{document}

\maketitle
\thispagestyle{firstpage}
\begin{abstract}
Co-existing wireless systems, which share a common spectrum, need to mitigate out-of-band (\OOB) radiation to avoid excessive interference.  For legacy systems, \OOB radiation is well understood and is commonly handled by digital precompensation techniques.  When using large arrays, however, new phenomena and hardware limitations have to be considered.  First, signals can be radiated directionally, which might focus the \OOB radiation.  Second, low-complexity hardware is used for cost reasons, which increases the relative amount of \OOB radiation.  Given that massive \MIMO and millimeter wave communication rely on base stations with a large number of antennas, the spatial behavior of \OOB radiation from large arrays will have significant implications for the hardware requirements of future base stations.  We show that, if the \OOB radiation is beamformed, its array gain is never larger than that of the in-band signal.  In many cases, the \OOB radiation is close to isotropic even when the in-band signal is highly directive.  With the same total radiated power, the \OOB radiation from large arrays is therefore never more severe than from a legacy system with the same adjacent-channel-leakage ratio.  Further, the \OOB radiation is less detrimental than from a legacy system since the high array gain of the in-band signal allows large arrays to radiate less total power than legacy systems.  We also show how \OOB radiation from large arrays varies with location in static propagation environments and how these effects vanish when averaged over the small-scale fading.  Since a higher relative amount of \OOB radiation can be tolerated for large arrays, the linearity requirement can be relaxed as compared to legacy systems.  Specifically, less stringent linearity requirements on each transmitter makes it possible to build large arrays from low-complexity hardware.
\end{abstract}

\section{Background}
\begin{figure}
	\centering
	\includegraphics[width=0.7\linewidth]{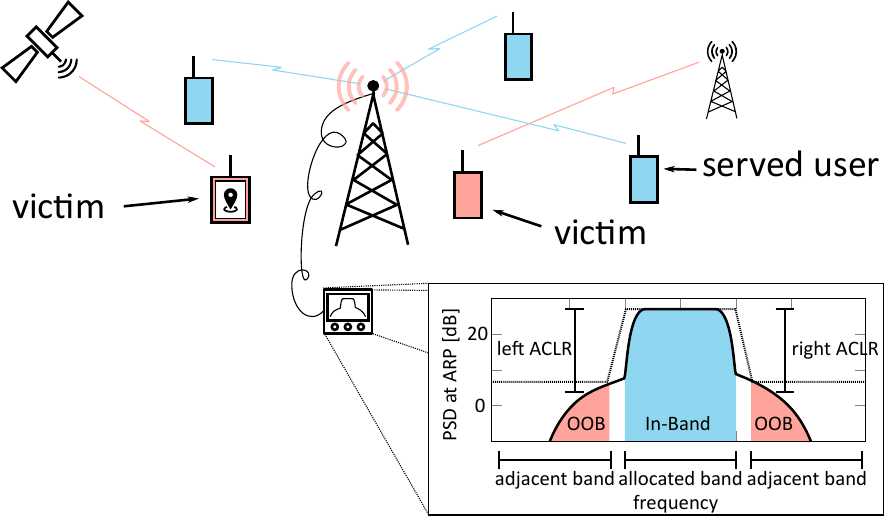}
	\caption{Victims of the \OOB radiation from the studied base station are other wireless systems operating in the vicinity.  To mitigate interference, hardware and algorithms for signal compensation are calibrated based on conducted measurements of the \ACLR at the antenna reference point.  A zoomed in sketch of the power spectral density of the transmitted signal is shown at the bottom.}
	\label{fig:scenario}
\end{figure}

Nonlinear hardware causes a radio system to emit spurious power outside its allocated frequency band.  This \textit{out-of-band (\OOB) radiation} is illustrated in Figure~\ref{fig:scenario}, which shows the power spectral density of a typical transmit signal.  The power outside the allocated band could harm the operation of a victim wireless system by interfering with its signal.  Therefore, the amount of \OOB radiation a transmitter is allowed to emit is regulated.  

The \emph{victim} of the \OOB radiation can be a system with a completely different application and sensitivity from the studied system---e.g., radar stations, telescopes for space research, \textsc{gnss} receivers, radio altimeters.  A victim should be distinguished from \emph{served users}, which are the receivers that operate within the allocated band and to whom the signal is intended.  The scenario is depicted in Figure~\ref{fig:scenario}, assuming a cellular system.

Commonly, standards require that \emph{conducted measurements}, i.e.\ measurements through a physical connection before the antenna at the \emph{antenna reference point} (\textsc{arp}), of the \ACLR (Adjacent-Channel Leakage Ratio) be below a certain threshold.  The \ACLR is the ratio between the power in the allocated band and the power in the strongest of the two adjacent bands.  The bandwidths of all bands are the same and the allocated band is centered around the carrier frequency and contains the whole desired signal including its excess bandwidth.  When there are multiple antennas, each power is measured as the total power summed over all antennas.

The goal of enforcing a constraint is to limit the absolute amount of interference that disturbs a victim.  Since \ACLR only measures this indirectly, an alternative measure is so called over-the-air measurements, where the actual \emph{received} \OOB power is measured.  This is treated further in Section~\ref{sec:measure}. 

\section{OOB Radiation from Large Arrays is Different}\label{sec:large_arrays}
Large arrays are envisioned to be used in both massive \MIMO and millimeter wave (mmWave) communication---both key technologies for 5G \cite{boccardi2014five}---to \emph{beamform} multiple signals to different users.  Constructive and destructive interference is used to limit overhearing between users and to increase the received signal power.  The extra power the user receives compared to if the signal were sent with the same total power from only one of the antennas is the \textit{array gain}. It can be used to either lower the radiated power compared to legacy systems or to increase the signal-to-interference-and-noise ratio of the received signal.  In many cases, especially in multicell scenarios, the system is interference limited, i.e.\ while it is possible to lower the transmit power substantially, the signal-to-interference-and-noise ratio can only be improved slightly \cite{marzetta2016fundamentals}.  The array gain is therefore assumed to be used to lower the transmit power.

The main difference between \OOB radiation from legacy \SISO systems and from large arrays is its \textit{spatial characteristics}, the amount of received power relative to the transmitted power at different points in space.  For legacy systems, \OOB radiation has the same spatial characteristics as the in-band signal.  For large arrays however, where the signal envelope and thus the nonlinear distortion is different at each antenna, the spatial characteristics of the \OOB radiation may differ from that of the in-band signal.  If the array gain of the \OOB radiation is small compared to that of the in-band signal, the low radiated power from large arrays means that a victim receives less \OOB power than from a legacy system with the same \ACLR requirement. 

Recently, \OOB radiation from large arrays has been studied in a large number of contributions to 3GPP.  Unlike the \OOB radiation from legacy systems \cite{gard1999characterization, ng2009coexistence}, however, \OOB radiation from large arrays has received little attention in the academic literature.  Models for phased arrays for satellite communication have been studied in, e.g., \cite{sandrin1973spatial}, phased arrays with two beams in \cite{hemmi2002pattern}.  Numerical results for large arrays in a frequency-flat system are presented in \cite{zou2015impact}.  In this paper, we use the analytical tools developed in \cite{mollen2016OOB} for frequency-selective systems to explain the spatial behavior of \OOB radiation and to discuss the implications for system design.   

The transmission from a large array is studied, e.g.\ the downlink in a cellular mobile system.  It is possible that a victim equipped with a large array could reject \OOB radiation by performing directive reception.  This possibility, however, is left for future research.  We illustrate the spatial characteristics of \OOB radiation for a system that serves multiple users by spatial multiplexing with nonlinear hardware.  The nonlinearity used to generate the illustrations is modeled by a third-order polynomial, whose coefficients are fitted to measured data from a class~\textsc{ab} amplifier.  The setup is agnostic to bandwidth and carrier frequency.  We consider signals with Gaussian amplitude distributions and high peak-to-average ratio (\PAR).  This includes most types of transmit signals, such as single-carrier and \OFDM signals, because the signals are precoded and are a combination of many independent symbols.  There are low-\PAR precoders that produce signals with lower amplitude variations, e.g.\ \cite{6565529}, to allow for less linear hardware.  While there is no reason to believe that the spatial characteristics of the \OOB radiation from such signals are any different, this remains to be shown in future research.  

The properties of the channel change with the carrier frequency.  At the high frequencies considered for mmWave communication, the channel has low rank, i.e.\ there are only one or a few distinct incident paths; whereas at lower frequencies there is more multipath propagation and isotropic scattering.  Measurements \cite{mammoetD12MaMiChannel, mmMagicD21ChannelMeasurements} reveal that, in reality, the channel has both low-rank and isotropic components and the relative significance of the two components changes continuously as frequency changes.  To reflect both mmWave and massive \MIMO communication, we will look at both a low-rank line-of-sight channel and a channel with isotropic scattering.  Both static and mobile propagation environments will be considered; see Table~\ref{tab:cases}.  

In line-of-sight communication, the time a mobile user equipment spends in one static lobe is relatively long.  For example, with 100 antennas separated by half a wavelength, the beamwidth is on the order 1.8$^\circ$ and a victim located \unit[100]{m} from the transmitter and moving at \unit[30]{m/s} perpendicular to the beam is inside the beam for \unit[100]{ms}.  We therefore model the line-of-sight channel as static, even if there is mobility.  

In an environment with isotropic scattering, the victim only has to move half a wavelength to experience a different channel.  The static and mobile scenarios therefore have to be studied separately.  In some static scenarios, the directivity of the \OOB radiation must be considered.  When either the served users are mobile or the victim is mobile, the amount of received \OOB radiation will change rapidly.  By coherent integration over several coherence times for example, a victim can protect its operation from outage in individual coherence times.  Therefore only the average \OOB radiation is relevant in mobile scenarios.

\begin{table}\centering\small
	\newlength{\mellanrum}\setlength{\mellanrum}{1.3ex}
	\newlength{\cellwidth}\setlength{\cellwidth}{6.5cm}
	\newcommand{\LOS}{%
		\parbox{2\cellwidth}{\raggedright
			\vspace{\mellanrum}%
		\begin{itemize}[leftmargin=*, noitemsep, nolistsep]
			\item beampattern is constant for long periods of time, both in static and mobile applications
			\item outage rate/risk matters
			\item beamforming of \OOB radiation has to be considered
		\end{itemize}
		\vspace{1ex}}
		}
	\newcommand{\NLOSstatic}{%
		\parbox{0.9\cellwidth}{\raggedright%
			\vspace{\mellanrum}%
		\begin{itemize}[leftmargin=*, noitemsep, nolistsep]
			\item coherence times is long compared\newline
			to the sensitivity of the victims
			\item outage rate/risk matters
			\item the “beampattern” is static
			\item beamforming of \OOB radiation has to be considered
		\end{itemize}
		\vspace{1ex}}%
		}
	\newcommand{\NLOSmobile}{%
		\parbox{1.1\cellwidth}{\raggedright%
			\vspace{\mellanrum}%
		\begin{itemize}[leftmargin=*, noitemsep, nolistsep]
			\item short coherence time
			\item ergodic rate/performance matters
			\item victims can code over many coherence times
			\item \OOB radiation beampattern changes rapidly
			\item average \OOB radiation has to be considered
		\end{itemize}
		\vspace{1ex}}%
		}
	\definecolor{LOScolor}{RGB}{0,185,231}
	\definecolor{NLOScolor}{RGB}{0,207,181}
	\definecolor{staticNLOScolor}{RGB}{23,199,210}
	\begin{tabular}{rp{0.9\cellwidth}p{1.1\cellwidth}}
		& \underline{static environment} & \underline{mobile environment}\\
		\rotatebox[origin=c]{90}{\underline{\smash{line-of-sight}}}&\multicolumn{2}{l}{\cellcolor{LOScolor!20}\LOS}\\
		\rotatebox[origin=c]{90}{\underline{\smash{isotropic scattering}}}&\cellcolor{staticNLOScolor!30}\NLOSstatic & \cellcolor{NLOScolor!25}\NLOSmobile\\
	\end{tabular}
	\caption{studied channels}
	\label{tab:cases}
\end{table}

\section{Line-of-Sight Channels}\label{sec:LOS}
\begin{figure}
	\centering
	\vspace{-2cm}
	\newsavebox{\illustrationbox}\savebox{\illustrationbox}{\includegraphics[scale=1]{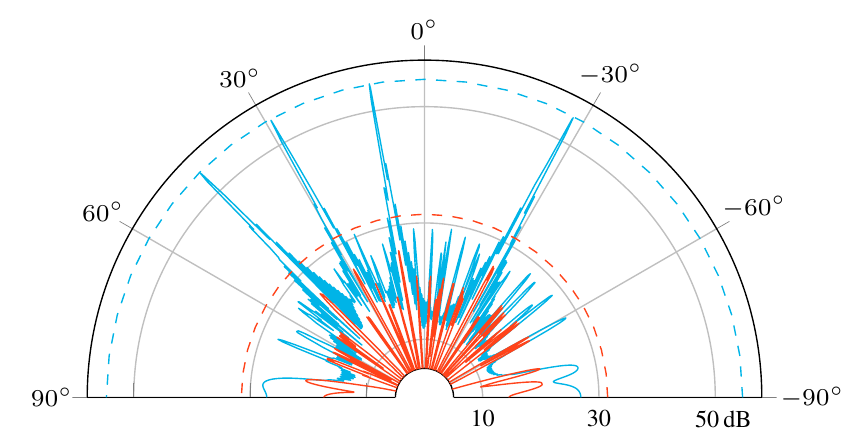}}
	\newlength{\illustrationlength}\settowidth{\illustrationlength}{\usebox{\illustrationbox}}
\setlength\tabcolsep{0pt}
	\hfill\clap{%
		\begin{tabular}{p{\illustrationlength}p{\illustrationlength}}
			\mbox{\hspace{1.01\illustrationlength}\llap{\includegraphics[scale=1]{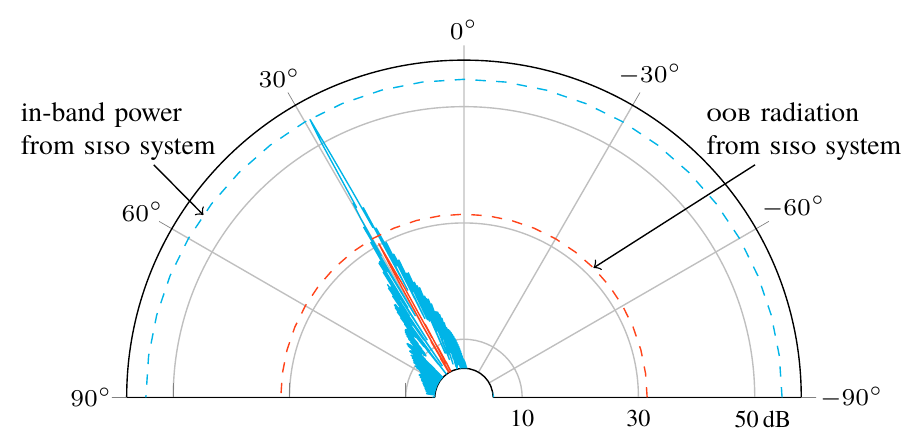}}}\newline
			\null\hfill 1 user, $P_\text{array} = \frac{1}{300} P_\text{\SISO}$\hfill\null
			&
			\usebox{\illustrationbox}\newline
			\null\hfill 4 users, $P_\text{array} = \frac{1}{75} P_\text{\SISO}$\hfill\null
			\\[4ex]
			\includegraphics[scale=1]{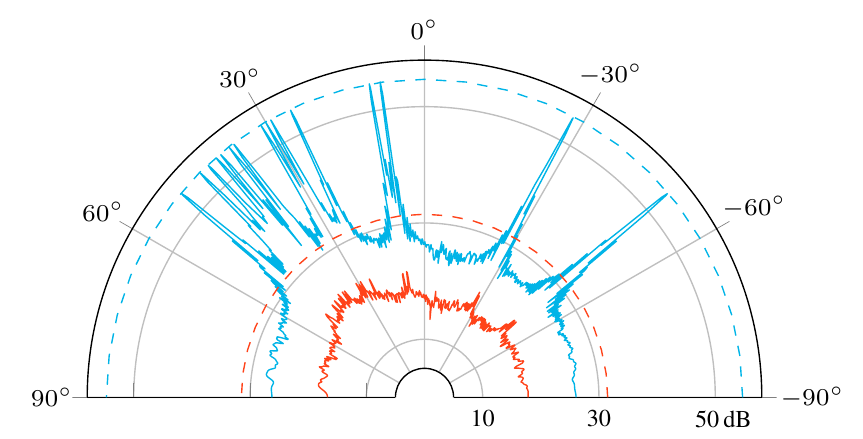}\newline
			\null\hfill 15 users, $P_\text{array} = \frac{1}{20} P_\text{\SISO}$\hfill\null
			&
			\includegraphics[scale=1]{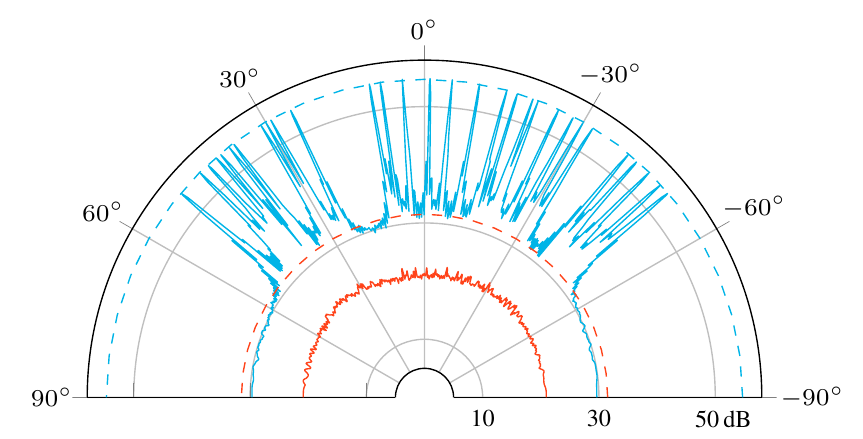}\newline
			\null\hfill 30 users, $P_\text{array} = \frac{1}{10} P_\text{\SISO}$\hfill\null
			\\[4ex]
			\includegraphics[scale=1]{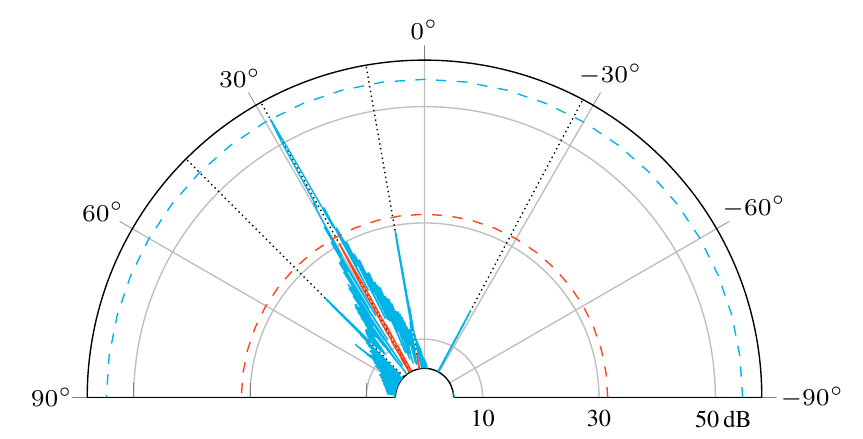}\newline
			\null\hfill 4 users distributed as in the sketch \rlap{$\longrightarrow$}\hfill\null
			&
			\null\hfill\smash{\includegraphics[]{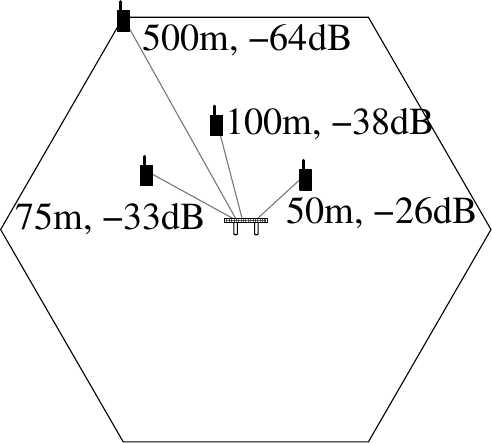}}\hfill\null
	\end{tabular}}\hfill\null

	\caption{The radiated in-band (solid blue) and out-of-band power (solid red) from a large uniform linear array, which has 300 antennas spaced by half a wavelength, that serves 1, 4, 15, 30 users through maximum-ratio precoding in line-of-sight.  The baud rate is \unit[20]{MHz} and the transmit signal has been pulse-shaped by a root-raised cosine with roll-off 0.22.  No narrowband assumption is made.  For comparison the radiated power from an isotropic \SISO system with the same \ACLR (\unit[23]{dB}) as the array is also shown (dashed lines).  At the bottom is a system with beams of different powers.  The beam power was chosen inversely proportional to the path loss, which is shown in the hexagonal sketch at the bottom right.  The transmitted power from the array $P_\text{array}$ and from the \SISO system $P_\text{\SISO}$ are scaled such that all users in the different systems receive the same power.  Source code is available \cite{mollen2016OOBcode}.}
	\label{fig:radiation_pattern_LOS}
\end{figure}

Figure~\ref{fig:radiation_pattern_LOS} illustrates how the in-band and \OOB beampatterns from a large array differ in line-of-sight.  It also shows how they compare to an omnidirectional \SISO system, whose transmit signal has the same \ACLR as the transmit signals of the array, and whose transmit power is chosen such that all users receive the same in-band power.  We see that there are bad directions, in which the \OOB radiation is stronger, and that there can be good directions, in which there is very little \OOB radiation.  

Note that the radiation pattern in a multi-user case can be similar to the single-user case with only one visible beam.  This happens when users experience very different path losses, which is common in a cellular system as shown in Figure~2.  Then most of the radiated power can be directed towards the weakest user, in order to give the same quality of service to all users.  

When the array serves one user in line-of-sight (one spatial component), the signal envelope is the same at each antenna, which creates distortion that has the same beamforming as the in-band signal.  There is therefore one direction, towards the served user, with \OOB radiation as bad as the \SISO system.  All other directions are suppressed in the same way as the in-band signal.  If the array serves multiple users (multiple spatial components), there are many bad directions.  All of the bad directions are better than in the \SISO system however, where the \OOB radiation in any direction always is greater than or equal to the \OOB radiation of the array in the worst direction.  When the number of served users increases, the bad and good directions disappear---all directions become equally good---and the \OOB radiation is significantly lower than in the \SISO system.

Just as the frequencies of the intermodulation products of a nonlinearity are combinations of the frequencies of the constituent components of the input signals, the spatial characteristics (the angle of departure in line-of-sight) of these products are combinations of the spatial characteristics of the input components.  When the number of “spatial intermodulation products”, which can be shown to grow super-linearly in the number of spatial components in the input signal (number of users times the number of channel taps), grows greater than the spatial dimension (i.e.\ the number of antennas), then the distortion becomes increasingly isotropic.

The array in Figure~\ref{fig:radiation_pattern_LOS} is a linear array with uniform antenna spacing of half a wavelength.  It creates radiation patterns without significant grating lobes, except for the back lobes on the opposite side of the array.  Other array geometries with grating lobes would cause the \OOB radiation to also radiate in the directions corresponding to those lobes.  Since arrays with grating lobes in general also have narrower beams, the probability that a victim ends up in a beam of \OOB radiation is not significantly changed by different array geometries.  Furthermore, the \OOB radiation in the directions of the grating lobes is still smaller than in the \SISO system.  The radiation patterns studied in Figure~\ref{fig:radiation_pattern_LOS} have the same basic appearance for any array type.

It is important to note that the array has no directions with worse \OOB radiation than the \SISO system.  Since the in-band signal is beamformed to maximize its array gain, the \OOB radiation can at most obtain the same array gain as the in-band signal, and therefore the \OOB radiation is never stronger than in a \SISO system with the same \ACLR.  

There is a small risk that a victim stands in a bad direction, especially if few users are served.  The worst case is when a system serves one user and a victim stands in the same direction as that user.  In this case, the victim receives as much \OOB radiation as from the \SISO system.  The probability that an unfortunate victim stands in a bad direction becomes smaller as the number of antennas grows large, and the main lobe becomes increasingly narrow. 

Based on this discussion, we make the following observations:
\begin{itemize}
	\item[(i)] Keeping the same \ACLR requirements as in legacy systems would guarantee that no victim, not even the most unfortunate one, receives more \OOB power than from a legacy system.  The \ACLR requirement does not have to be more stringent.
	
	\item[(ii)] If multiple users are served, the \OOB radiation can be treated as isotropic and the legacy \ACLR requirement can be relaxed.
	
	\item[(iii)] If a single user is served, the \OOB radiation is highly directive and the legacy \ACLR requirement can be relaxed if a certain probability is allowed that an unfortunate victim ends up in an \OOB lobe.  This probability is increasingly small when the array is large.
\end{itemize}

\section{Static Channels with Isotropic Fading}
The received \OOB radiation varies with the channels to the victim and the served users.  If the channel changes slowly or if the victim systems is sensitive to outage, e.g.\ when there are high reliability or latency requirements, the \OOB radiation has to be constrained during every channel realization.  Much of what was said in Section~\ref{sec:LOS} about static line-of-sight channels carries over to slowly changing frequency-selective channels with isotropic fading.  One difference, however, comes from the larger number of propagation paths.  

A consequence of the multipath propagation of wideband signals is frequency-selective fading.  The multiple taps of the channel make the \OOB radiation less directive in much the same way as serving more users.  Therefore, also when a single user is served by the array, the most unfortunate victim of \OOB radiation still receives much less power than from a \SISO system.  

\begin{figure}
	\centering
	\definecolor{linecolor}{rgb}{.5,.5,.5}
	\begin{tikzpicture}
		\begin{semilogyaxis}[
		xlabel = {received \OOB power at victim (rel.\ to the thermal noise power) [dB]},
		ylabel = {\textsc{ccdf}},
		ymax = 1,
		ymin = 10^(-4),
		xmax= -9,
		xmin= -42,
		width = \linewidth,
		height= 38ex,
		legend cell align=left,
		legend style={legend pos=south west, font=\scriptsize},
		]
		\draw[color=linecolor, line width=0.4pt, opacity=0.5] (axis cs:-14.685210829,0.00001) -- (axis cs:-14.685210829,1);
		\node [rotate=0, align=left, anchor=west] at (axis cs: -18.5, 10^-2) {\SISO system,\\ tx power ${=}P_\text{\SISO}$};
		\addplot[color=color3, line width = 1pt] table [col sep = comma] {./illustrationer/OOB_CCDF_dB_M1K1.csv};

		\draw[color=linecolor, line width=0.4pt, opacity=0.5] (axis cs:-34.6654571032917,0.00001) -- (axis cs:-34.6654571032917,1);
		\node [rotate=0, align=left, anchor=west] at (axis cs: -39.85, 10^-2){large array,\\ 100 antennas,\\ 1 user,\\tx power ${=}\frac{P_\text{\SISO}}{100}$};
		\addplot[color=color3, line width = 1pt] table [col sep = comma] {./illustrationer/OOB_CCDF_dB_M100K1.csv};

		\draw[color=linecolor, line width=0.4pt, opacity=0.5] (axis cs:-24.6852108295,0.00001) -- (axis cs:-24.6852108295,1);
		\node [rotate=0, align=left, anchor=west] at (axis cs: -31.1, 10^-2){large array, \\100 antennas, \\10 users,\\tx power ${=}\frac{P_\text{\SISO}}{10}$ \\(equal power \\allocation)};
		\addplot[color=color3, line width = 1pt] table [col sep = comma] {./illustrationer/OOB_CCDF_dB_M100K10.csv};
		\end{semilogyaxis}
	\end{tikzpicture}
	\caption{The distribution of the power received by a victim in the adjacent band in an \IID Rayleigh fading propagation environment with a delay spread equal to 15 symbol periods.  The radiated power is normalized such that the served users receive the same amount of in-band power (same signal-to-noise ratio) as they would have if they were served one-by-one by a \SISO system with transmit power $P_\text{\SISO}$.  The transmit signals have the same \ACLR in all cases.  The signals to each of the ten users are assumed to be equally strong and, likewise, all path losses are equal.  The mean received power is marked by vertical lines.  Source code is available \cite{mollen2016OOBcode}.}
	\label{fig:channel_gain_CCDF}
\end{figure}
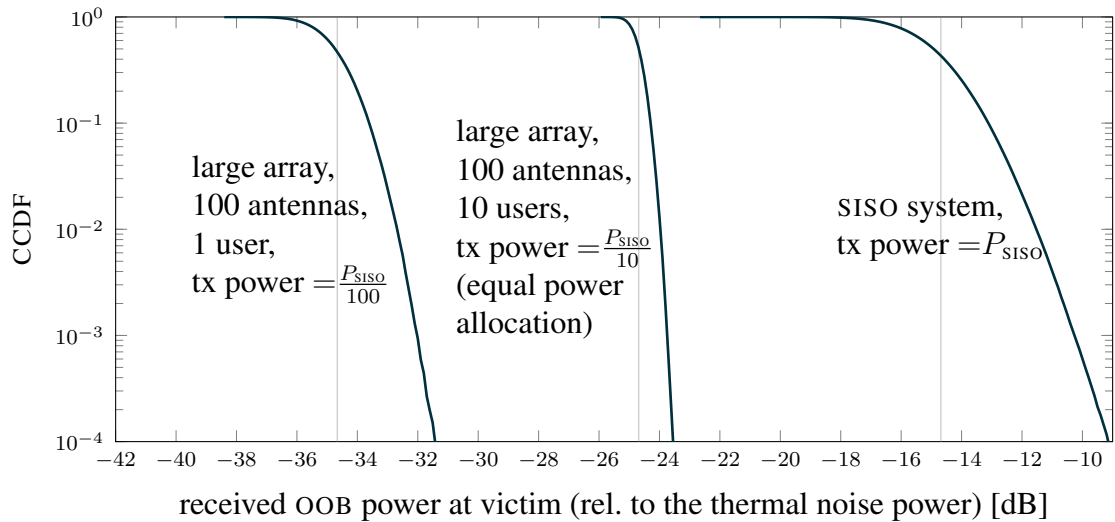

Another advantage of the large array as compared to the \SISO system is channel hardening.  Figure~\ref{fig:channel_gain_CCDF} shows the distribution of the received \OOB power at a random victim for different systems with transmit signals with the same \ACLR.  It can be seen how constructive and destructive superposition, which is the result of multipath propagation, can result in large variations in \OOB radiation in the \SISO system.  In a large array, however, channel hardening eliminates variations due to multipath propagation; variations only come from the directivity of the transmission.  Just like in the line-of-sight system, the \OOB radiation of the single-user system is slightly directive and there is a small risk that a victim will receive more \OOB radiation than on average---in Figure~\ref{fig:channel_gain_CCDF}, the probability to receive \unit[3]{dB} more \OOB radiation than on average is 0.001.  The directivity becomes less prominent when the number of significant users, i.e.\ users to whom a significant part of the transmit power is directed, is increased.  With ten equally significant users, the vertical slope in Figure~\ref{fig:channel_gain_CCDF} shows that the \OOB radiation is practically isotropic.  As noted in Section~\ref{sec:LOS}, however, large differences in path loss between the served users can require that most of the power is beamformed to a single user if all users are to enjoy the same quality of service.  The single-user case is therefore representative also for many multi-user systems.

\begin{figure}
	\vspace{-1cm}
\centering
	map\\
	\includegraphics[height=0.26\textheight]{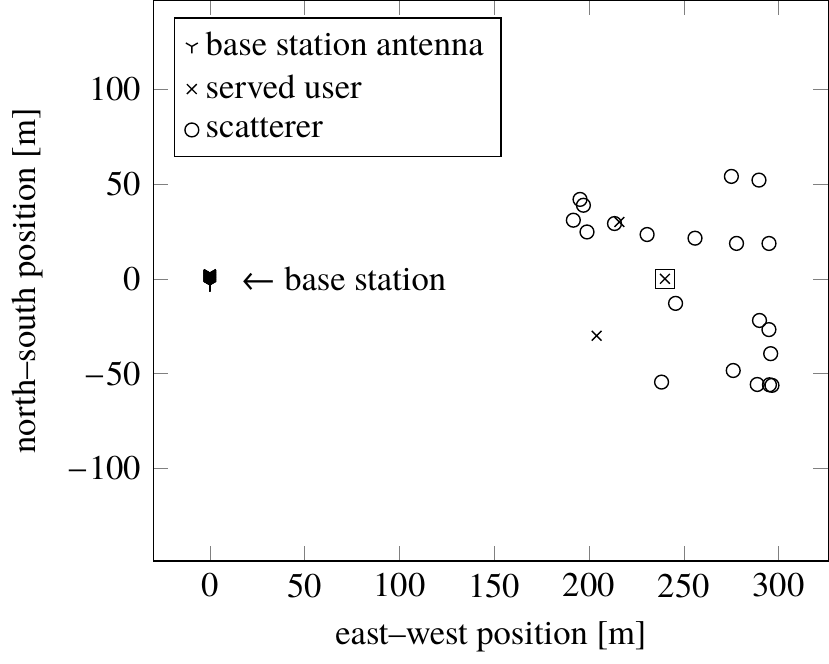}\\
	in-band power\\
	\mbox{\llap{{\includegraphics[]{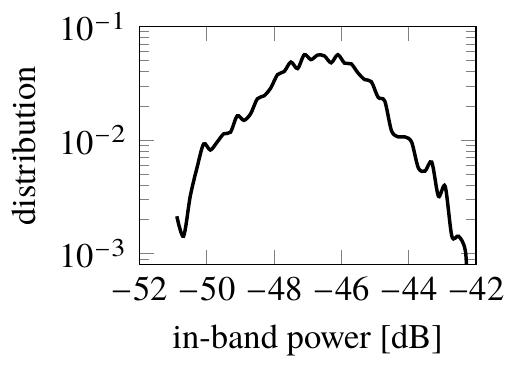}}}\includegraphics[height=0.26\textheight]{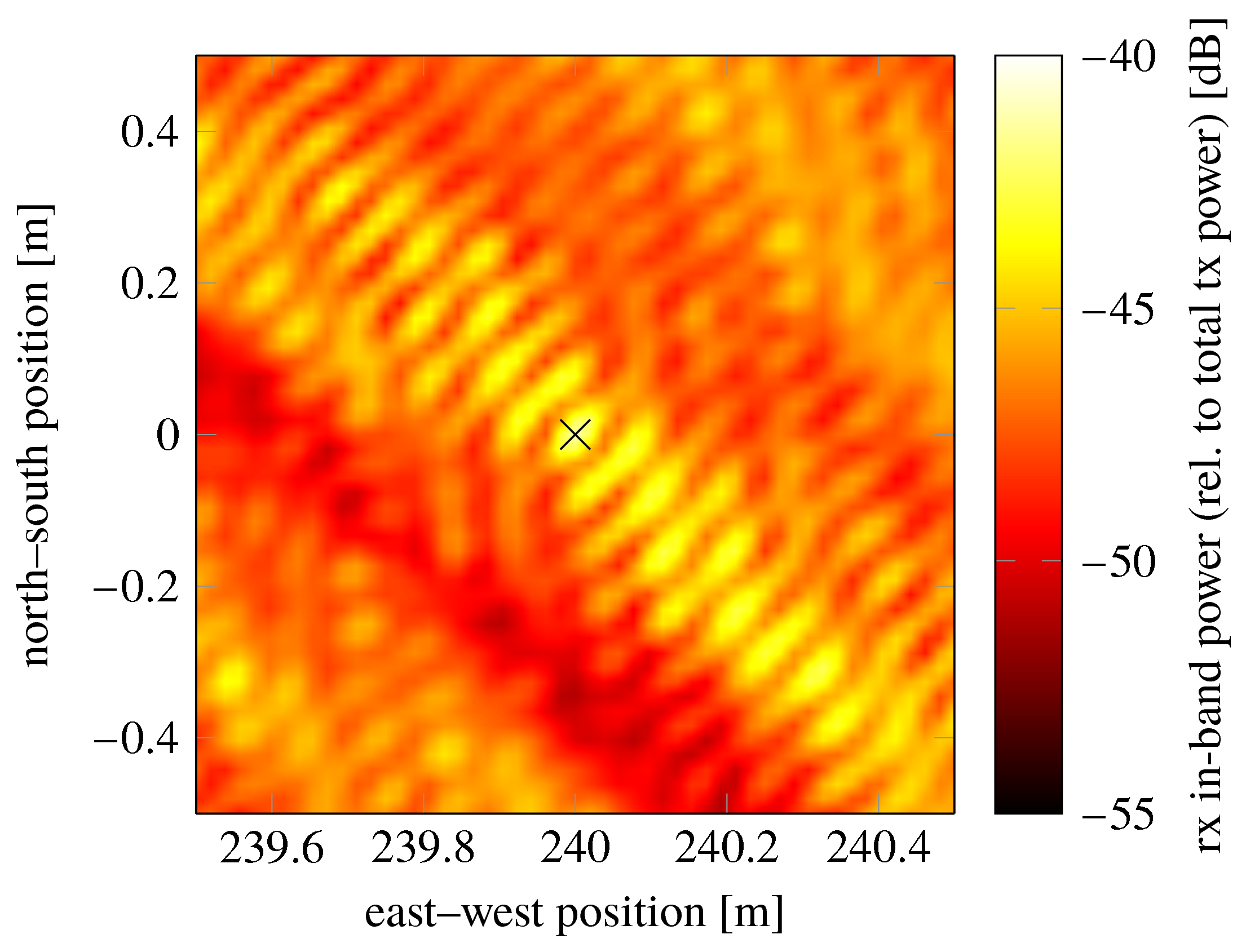}}\\
	\OOB power\\
	\mbox{\llap{{\includegraphics[]{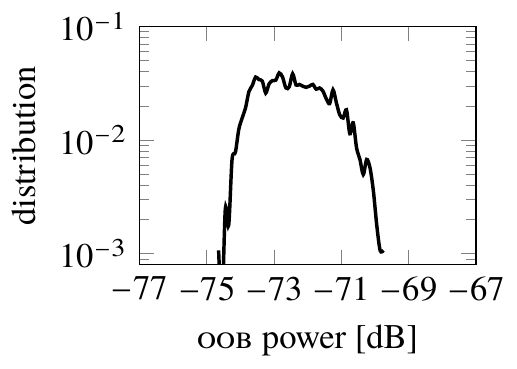}}}\includegraphics[height=0.26\textheight]{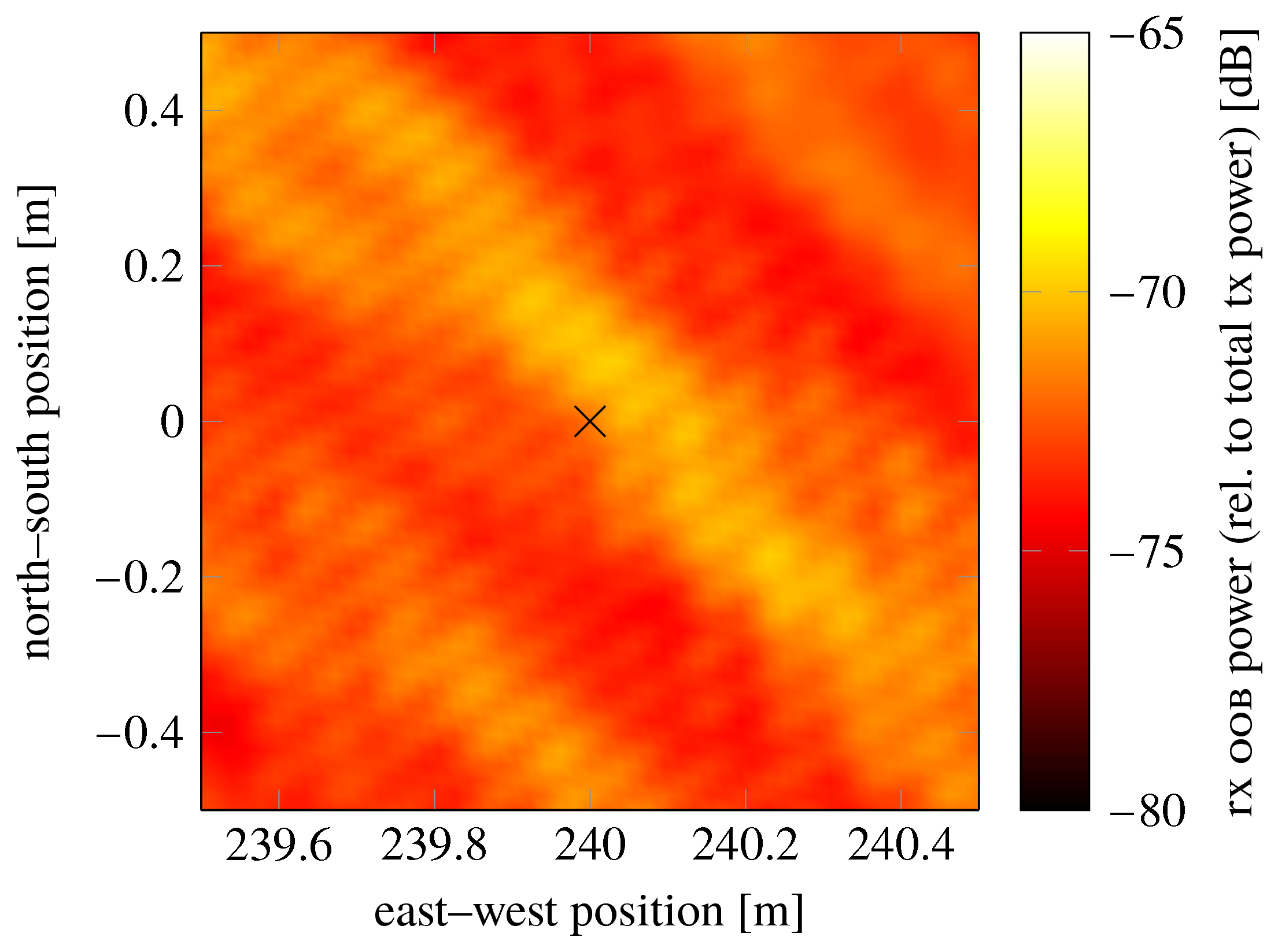}}
	
\caption[\null]{Heat map of in-band and \OOB signal power intensity over the area marked $\square$ in the uppermost figure, where the geometry of the setup can be seen.  A linear uniform array with 100 antennas, half a wavelength apart, is located at the origin and 20 scatterers and three users are randomly placed in a \unit[100]{m} large quadratic area \unit[250]{m} east of the array.  The empirical distribution of the received power over the shown area is given to the left of the figures.
Source code is available \cite{mollen2016OOBcode}.}
\label{fig:karta}
\end{figure}

In Figure~\ref{fig:karta}, a simple scattering environment is illustrated.  Scattering centers have been randomly dropped over an area and a uniform linear array with 100 antennas beamforms to three users inside the area.  It can be seen that the directivity, or the array gain, of both the in-band and the \OOB signal varies with location.  The variations, however, are much smaller for the \OOB signal because of its isotropy and because of channel hardening.

\section{Mobile Channels with Isotropic Fading}
When the disturbing \OOB power can be averaged over many channel fades and all gains can be averaged over the fading, the victim can protect its operation from outage as long as the average received \OOB power is limited.  In a mobile channel with isotropic fading, we have to distinguish between two cases:
\begin{description}
	\item[Case~1] the channels to the served user and to the victim are uncorrelated,
	
	\item[Case~2] the channels are correlated.
\end{description}
Whereas Case~1 is the common one, Case~2 is perceivable when the served user and victim are served by different transmitters but share the same antenna and their channel is not frequency selective enough to decorrelate the channels of the two bands.

In Case~1, the average \OOB power that the victim receives, normalized by the path loss, is determined by the total radiated \OOB power at the transmitter, both for legacy systems and for large arrays.  Given a transmit power, an \ACLR constraint thus limits how much \OOB power that a victim receives on average, which is enough to protect the operation of the victim.  When the correlation is high, the \OOB radiation of Case~2 has to be analyzed as in the static case, since the \OOB radiation to the victim then experiences an array gain also when averaged over many fades.

Since less radiated power is required from a large array than from a legacy system for a given received in-band power, the average received \OOB power is also correspondingly lower when the transmit signals of two systems have the same \ACLR, which was seen in Figure~\ref{fig:channel_gain_CCDF}.  The \ACLR requirement should therefore be relaxed for the large array as compared to the legacy system by the same amount, by which the total radiated power is reduced.  Since the in-band array gain grows with the number of antennas of the array, the \ACLR requirement can be relaxed more, the more antennas the array has.  However, the radiated power also increases with the number of served users and varies slightly depending on the employed beamforming technique.  Therefore the \ACLR requirement has to be specified in terms of these system parameters or set according to the worst scenario, in which the most \OOB power is radiated.  

\begin{figure}
	\centering
	\includegraphics[width=0.85\linewidth]{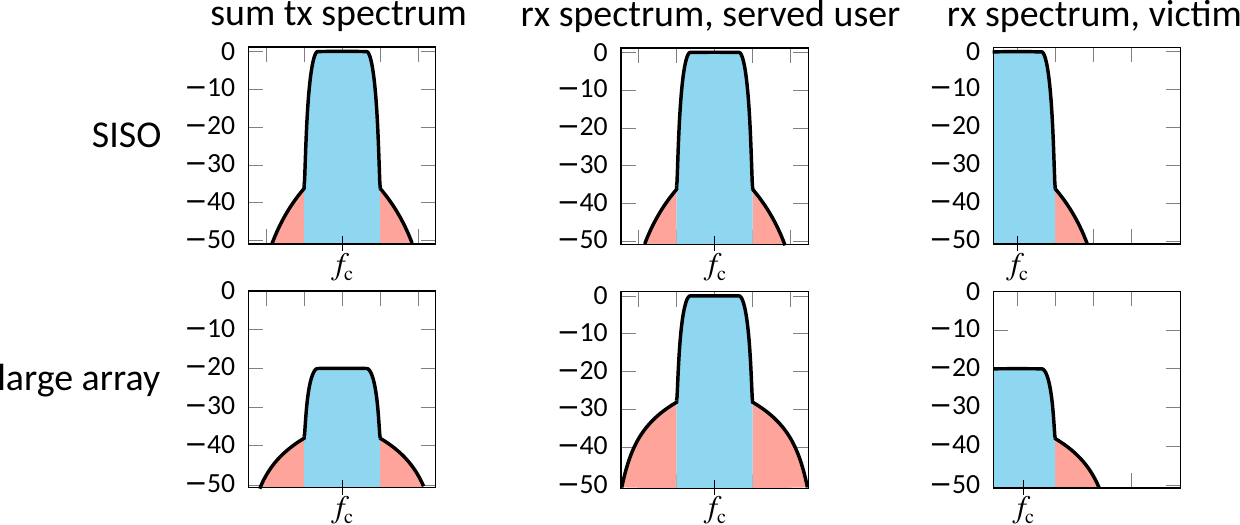}
	\caption[\null]{average power spectral densities (in dB relative to received power of the served user) in a \SISO system and in a system with a large array}
	\label{fig:psd_comp}
\end{figure}

Figure~\ref{fig:psd_comp} shows the average power spectral densities of two example scenarios; the path loss has been normalized for simplicity.  In the legacy \SISO case, highly linear hardware gives the transmitted signal a good \ACLR.  Consequently, the served user receives a sufficient amount of in-band power and, at the same time, the victim who operates in the adjacent band receives little disturbing power.  In the large array case, the transmitted signal has an \ACLR that is seemingly worse because less linear hardware is used; the transmitted power is also smaller.  Because the signal is beamformed, however, the served user still receives a sufficient amount of in-band power.  At the same time, the victim receives little disturbing power on average.  This example shows that the \ACLR constraint of \SISO systems cannot directly be applied to arrays.  The array gain of the in-band signal at the served users and the distribution of the \OOB signal at the victim also have to be taken into account.  

\section{How to Measure OOB Radiation}\label{sec:measure}
To mitigate the disturbance of other systems, most communication standards, such as \textsc{wcdma}, \textsc{lte}, \textsc{w}i\textsc{f}i, and national regulatory bodies, such as FCC (the Federal Communications Commission) in the United States, limit the amount of permitted \OOB radiation.  This is usually done by enforcing a constraint on the \ACLR of the transmit signal and a maximum power level for the emitted \OOB signal.  For example, in \textsc{lte}, the \ACLR has to be better than \unit[45]{dBc} or the absolute power spectral density of the signal in a wide outdoor area has to be lower than \unit[$-$13]{dBm/MHz} outside the allocated band, whichever is less stringent.

Two quantities are of interest: the useful in-band power at the served users, and the disturbing \OOB power at the victim.  Under the assumption that both powers are attenuated equally much, the \ACLR of the transmitted signal is the ratio between the two.  When the transmitter has a large array, the array gain will influence the received powers.  We have seen that, when the powers can be averaged over many coherence intervals, the disturbing \OOB radiation is isotropic, while the useful in-band signal gets a large array gain.  The \ACLR regulations used in legacy systems that do not consider array gain are therefore unnecessarily stringent, and can be relaxed for large arrays by an amount equal to the array gain of the in-band signal.

In static scenarios, a constraint might have to include a safety margin to protect sensitive victims from the case when the array gain of the \OOB radiation is significant.  Because the \OOB array gain is smaller than the in-band array gain and the transmit power is lower than in legacy systems, the \ACLR constraint with added safety margin is still relaxed compared to legacy systems.  For isotropic scattering, the safety margin can be read off from percentiles, like the one in Figure~\ref{fig:channel_gain_CCDF}; it is often small and can be neglected however.  In a line-of-sight channel, this margin can be substantial and can be measured by the served users.

Since \OOB radiation is isotropic in many cases, a more practical way to put the \OOB constraint could be to regulate the total radiated \OOB power in relation to the in-band power.  FCC \cite[Sec.\ IV.G.3]{federal2016fcc} has also mentioned the possibility to measure \OOB radiation “over the air”, i.e.\ to take measurements at selected positions around the transmitting array and draw conclusions about the received \OOB power everywhere else from those measurements.  For practical reasons, over-the-air measurements is most likely the only alternative for mmWave arrays.

One way to do that, in analogy to the legacy \ACLR measure, would be to set up a controlled environment and let the array beamform a signal to a served user in its normal operational mode.  Then measure the useful in-band power at the served user and the disturbing \OOB power at a reference victim.  The ratio between the two---the \emph{array \ACLR}---can then be constrained in the same way as in legacy systems.  The reason for employing such a strategy would be to avoid measuring directly on each of the individual transmit signals in the array and to make the constraint independent of the number of antennas and other system parameters.  

If the same transmitted power is used as in legacy systems, the \OOB constraint cannot necessarily be relaxed as compared to legacy systems.  An stricter \OOB constraint, however, is only necessary if there is a non-negligible risk that the array gain of the disturbing \OOB radiation at a victim is large.  We have shown how this risk is increasingly small for large arrays.  Furthermore, as discussed in Section~\ref{sec:large_arrays}, to use the same transmit power as in a legacy system is seldom necessary.  

\section{Conclusion}
We have shown how a victim, on average, receives less \OOB radiation from base stations with large arrays than from legacy systems with the same \ACLR requirement for a given received \SNR requirement.  In the worst case, the victim receives the same amount of \OOB radiation as from the legacy system.  However, this worst case event occurs only when most of the transmitted power is directed towards a single user whose channel impulse response consists of taps that are all linearly dependent, and then only with a small probability.  Furthermore, the probability grows increasingly small as the number of antennas grows large.

This conclusion relies on the assumption that the channel to the victim is uncorrelated to the channel of the served users.  If that is not true, for example if the victim shares its antenna with one of the served users, the probability that the victim receives the same amount of \OOB radiation as from the legacy system can increase significantly.  However, the \OOB radiation is never greater than from the legacy system.

When the dimension of the space spanned by the channel vectors of the served users is large, which happens with high probability when the product of the number of users and number of channel taps is large, the \OOB radiation becomes close to isotropic.  This makes it redundant to measure the radiation pattern for each setup, which simplifies the measurement of \OOB radiation.

This suggests that relaxed \ACLR and linearity constraints can be used for the hardware in large arrays.  To set appropriate linearity requirements on the hardware is important, because it will be decisive for how future radio equipment will be designed.  Especially since it is desirable to build large arrays without high-end hardware or advanced compensation techniques, which become impediments as the number of radio chains grows \cite{UGUSGC14, bjornson2015massive}.  The linearity requirement will determine what amplifier architectures, digital-to-analog converters etc.\ that have to be employed.  It will also influence what signal processing is required, such as predistortion, \PAR reduction and low-\PAR precoding.  Power efficiency, system complexity, cost and size of future communication systems will all be affected by the way \OOB radiation from large arrays is regulated.  

\bibliographystyle{IEEEtran}
\bibliography{bib_forkort_namn,bibliografi}

\begin{thebibliography}{10}
\providecommand{\url}[1]{#1}
\csname url@samestyle\endcsname
\providecommand{\newblock}{\relax}
\providecommand{\bibinfo}[2]{#2}
\providecommand{\BIBentrySTDinterwordspacing}{\spaceskip=0pt\relax}
\providecommand{\BIBentryALTinterwordstretchfactor}{4}
\providecommand{\BIBentryALTinterwordspacing}{\spaceskip=\fontdimen2\font plus
\BIBentryALTinterwordstretchfactor\fontdimen3\font minus
  \fontdimen4\font\relax}
\providecommand{\BIBforeignlanguage}[2]{{%
\expandafter\ifx\csname l@#1\endcsname\relax
\typeout{** WARNING: IEEEtran.bst: No hyphenation pattern has been}%
\typeout{** loaded for the language `#1'. Using the pattern for}%
\typeout{** the default language instead.}%
\else
\language=\csname l@#1\endcsname
\fi
#2}}
\providecommand{\BIBdecl}{\relax}
\BIBdecl

\bibitem{boccardi2014five}
F.~Boccardi, R.~W. Heath, Jr., A.~Lozano, T.~L. Marzetta, and P.~Popovski,
  ``Five disruptive technology directions for {5G},'' \emph{{IEEE} Commun.\
  Mag.}, vol.~52, no.~2, pp. 74--80, Feb. 2014.

\bibitem{marzetta2016fundamentals}
T.~L. Marzetta, E.~G. Larsson, H.~Yang, and H.~Q. Ngo, \emph{Fundamentals of
  Massive {MIMO}}.\hskip 1em plus 0.5em minus 0.4em\relax Cambridge University
  Press, 2016.

\bibitem{gard1999characterization}
K.~G. Gard, H.~M. Gutierrez, and M.~B. Steer, ``Characterization of spectral
  regrowth in microwave amplifiers based on the nonlinear transformation of a
  complex {Gaussian} process,'' \emph{{IEEE} Trans.\ Microw.\ Theory Tech.},
  vol.~47, no.~7, pp. 1059--1069, Jul. 1999.

\bibitem{ng2009coexistence}
M.~H. Ng, S.-D. Lin, J.~Li, and S.~Tatesh, ``Coexistence studies for {3GPP}
  {LTE} with other mobile systems,'' \emph{{IEEE} Commun.\ Mag.}, vol.~47,
  no.~4, pp. 60--65, Apr. 2009.

\bibitem{sandrin1973spatial}
W.~Sandrin, ``Spatial distribution of intermodulation products in active phased
  array antennas,'' \emph{{IEEE} Trans.\ Antennas Propag.}, vol.~21, no.~6, pp.
  864--868, Nov. 1973.

\bibitem{hemmi2002pattern}
C.~Hemmi, ``Pattern characteristics of harmonic and intermodulation products in
  broadband active transmit arrays,'' \emph{{IEEE} Trans.\ Antennas Propag.},
  vol.~50, no.~6, pp. 858--865, Jun. 2002.

\bibitem{zou2015impact}
Y.~Zou, O.~Raeesi, L.~Antilla, A.~Hakkarainen, J.~Vieira, F.~Tufvesson, Q.~Cui,
  and M.~Valkama, ``Impact of power amplifier nonlinearities in multi-user
  massive {MIMO} downlink,'' in \emph{Proc.\ {IEEE} Globecom Workshops}, Dec.
  2015, pp. 1--7.

\bibitem{mollen2016OOB}
C.~Mollén, U.~Gustavsson, T.~Eriksson, and E.~G. Larsson, ``Out-of-band
  radiation measure for {MIMO} arrays with beamformed transmission,'' in
  \emph{Proc.\ {IEEE} Int.\ Conf.\ Commun.}, May 2016, pp. 1--6.

\bibitem{6565529}
S.~Mohammed and E.~G. Larsson, ``Constant-envelope multi-user precoding for
  frequency-selective massive {MIMO} systems,'' \emph{{IEEE} Wireless Commun.\
  Lett.}, vol.~2, no.~5, pp. 547--550, Oct. 2013.

\bibitem{mammoetD12MaMiChannel}
O.~Edfors and F.~Tufvesson, ``{MaMi} channel characteristics: {Measurement}
  results,'' \url{https://mammoet-project.eu/publications-deliverables},
  Massive {MIMO} for Efficient Transmission (MAMMOET), Project Deliverable
  D1.2, Jun. 2015, online: accessed 2016-10-26.

\bibitem{mmMagicD21ChannelMeasurements}
M.~Peter, K.~Sakaguchi, S.~Jaeckel, S.~Wu, M.~Nekovee, J.~Medbo, K.~Haneda,
  S.~L.~H. Nguyen, R.~Naderpour, J.~Vehmas, F.~Mani, A.~Bamba, R.~D’Errico,
  M.~Rybakowski, J.-M. Conrat, A.~Goulianos, P.~Cain, M.~Rumney, M.~Dieudonne,
  H.~Wang, and M.~Kottkamp, ``Measurement campaigns and initial channel models
  for preferred suitable frequency ranges,''
  \url{https://bscw.5g-mmmagic.eu/pub/bscw.cgi/93650}, Millimetre-Wave Based
  Mobile Radio Access Network for Fifth Generation Integrated Communications
  (mmMagic), Project Deliverable H2020-ICT-671650-mmMAGIC/D2.1, Mar. 2016,
  online: accesed 2016-10-26.

\bibitem{mollen2016OOBcode}
\BIBentryALTinterwordspacing
C.~Mollén. (2016, Nov.) Source code for the analysis of out-of-band radiation
  from large arrays. [Online]. Available:
  \url{https://github.com/OOBRadMIMO/NumericalResults}
\BIBentrySTDinterwordspacing

\bibitem{federal2016fcc}
``{FCC} 16-89 report and order and further notice of proposed rulemaking,''
  \url{https://www.fcc.gov/document/spectrum-frontiers-ro-and-fnprm}, Federal
  Communications Commission, Tech. Rep., Jul. 2016, online: accessed
  2016-10-26.

\bibitem{UGUSGC14}
U.~Gustavsson, C.~{Sanchez Perez}, T.~Eriksson, F.~Athley, G.~Durisi, P.~N.
  Landin, K.~Hausmair, C.~Fager, and L.~Svensson, ``On the impact of hardware
  impairments on massive {MIMO},'' in \emph{Proc.\ {IEEE} Global Commun.\
  Conf.}, Dec. 2014.

\bibitem{bjornson2015massive}
E.~Björnson, M.~Matthaiou, and M.~Debbah, ``Massive {MIMO} with non-ideal
  arbitrary arrays: {Hardware} scaling laws and circuit-aware design,''
  \emph{{IEEE} Trans.\ Wireless Commun.}, vol.~14, no.~8, pp. 4353--4368, Aug.
  2015.

\end{thebibliography}
\end{document}